\documentclass[onecollarge,natbib]{svjour2}
\bibpunct{[}{]}{;}{n}{}{,} 
\usepackage{graphicx}
\usepackage{amsmath}

\journalname{Few-Body Systems}
\begin{document}

\title{Transversity Form Factors and Generalized Parton Distributions of the pion in chiral quark models%
\thanks{Presented by W. Broniowski at LIGHTCONE 2011, 23 - 27 May, 2011, Dallas}
\thanks{Supported by the Bogoliubov-Infeld program (JINR), the Polish Ministry of
Science and Higher Education, grants N~N202~263438 and N~N202~249235,
Spanish DGI and FEDER grant FIS2008-01143/FIS, Junta de Andaluc{\'{\i}}a
grant FQM225-05, and EU Integrated Infrastructure Initiative Hadron Physics
Project, contract RII3-CT-2004-506078. AED acknowledges partial support from
the Russian Foundation for Basic Research, projects No.~10-02-00368 and No.~11-02-00112.}}

\titlerunning{Transversity form factors and GPD}        

\author{Wojciech Broniowski         \and
        Alexander E. Dorokhov       \and
        Enrique Ruiz Arriola
}

\authorrunning{W. Broniowski et al.} 

\institute{Wojciech Broniowski,  \email{Wojciech.Broniowski@ifj.edu.pl}  \at
              The H. Niewodnicza\'nski Institute of Nuclear Physics, Polish Academy of Sciences, PL-31342 Krak\'ow, Poland \\
              Institute of Physics, Jan Kochanowski University, PL-25406~Kielce, Poland
           \and
           Alexander E. Dorokhov, \email{dorokhov@theor.jinr.ru} \at
             Joint Institute for Nuclear Research, Bogoliubov Laboratory of Theoretical Physics, 141980, Dubna, Russia \\
             Institute for Theoretical Problems of Microphysics, Moscow State University, RU-119899, Moscow, Russia
           \and
           Enrique Ruiz Arriola,  \email{earriola@ugr.es}  \at
             Departamento de F\'{\i}sica At\'omica, Molecular y Nuclear, Universidad de Granada, E-18071 Granada, Spain \\
             Instituto Carlos I de Fisica Te\'orica y Computacional, Universidad de Granada, E-18071 Granada, Spain
}

\date{30 August 2011}

\maketitle

\begin{abstract}
The transversity Generalized Parton Distributions (tGPDs) and related
transversity form factors of the pion are evaluated in chiral quark
models, both local (Nambu--Jona-Lasinio) and nonlocal, involving a
momentum-dependent quark mass. The obtained tGPDs satisfy all a priori
formal requirements, such as the proper support, normalization, and
polynomiality. We evaluate generalized transversity form factors
accessible from the recent lattice QCD calculations. These form
factors, after the necessary QCD evolution, agree very well with the
lattice data, confirming the fact that the spontaneously broken chiral
symmetry governs the structure of the pion also in the case of the
transversity observables.  
\end{abstract}

\newpage

This presentation is based on our two recent papers
\cite{Broniowski:2010nt,Dorokhov:2011ew}, where more results and
details may be found.  The tGPDs are the least-explored Generalized
Parton Distributions (for extensive reviews
see~\cite{Belitsky:2005qn,Feldmann:2007zz,Boffi:2007yc} and references
therein). For the quark sector they correspond to the hadronic matrix
elements of the tensor quark bilinears, $\bar{q}(x) \sigma^{\mu \nu}
q(0)$, and are the maximum-helicity chirally odd objects.  Information
on these rather elusive entities comes from the recent lattice
determination \cite{Brommel:2007xd} of the pion transversity form
factors (tFFs), defined via moments of tGPDs in the Bjorken $x$
variable. That way the lattice calculations provide a direct path to
verify the underlying models for quantities which are very difficult
to be accessed experimentally.

Our analysis consist of two distinct parts: 1)~the model determination
of tFFs and tGPDs of the pion and 2)~the QCD evolution.  For the first
part and within a non perturbative setup we apply chiral quark models
which have proven to be very useful in the determination of the soft
matrix elements entering various high-energy processes
\cite{Davidson:1994uv,RuizArriola:2001rr,Davidson:2001cc,%
Broniowski:2003rp,Dorokhov:1998up,Polyakov:1999gs,Dorokhov:2000gu,Anikin:2000th,Praszalowicz:2002ct,%
Praszalowicz:2003pr,Bzdak:2003qe,Holt:2010vj,Nguyen:2011jy,Polyakov:1999gs,Theussl:2002xp,Bissey:2003yr,%
Noguera:2005cc,Broniowski:2007si,Frederico:2009pj,Frederico:2009fk,Esaibegian:1989uj,Dorokhov:1991nj,%
Petrov:1998kg,Anikin:1999cx,Praszalowicz:2001wy,Dorokhov:2002iu,RuizArriola:2002bp,RuizArriola:2002wr,Broniowski:2008hx,%
Tiburzi:2005nj,Broniowski:2007fs,Courtoy:2007vy,Courtoy:2008af,Kotko:2008gy}.
We use the standard local Nambu--Jona-Lasinio (NJL) model with the
Pauli-Villars regularization \cite{RuizArriola:1991gc} and two
versions of the nonlocal models, where the quark mass depends on the
virtuality: the instanton-motivated model \cite{Diakonov:1985eg} and
the Holdom-Terning-Verbeek (HTV) model \cite{Holdom:1990iq} (these
variants differ in the form of the quark-pion vertex).

The second ingredient is the QCD evolution where renormalization improved
radiative and perturbative gluonic corrections are added to the low
energy matrix element. The scale where the quark model calculation is
carried out can be identified with the help of the momentum fraction
carried by the quarks. According to phenomenological
extractions~\cite{Sutton:1991ay,Gluck:1999xe} as well as lattice
calculations~\cite{Best:1997qp}, the valence quarks carry about 50\%
of the total momentum at the scale $\mu= 2{\rm GeV}$.  In quark
models, having no explicit gluon degrees of freedom, the valence
quarks carry 100\% of the momentum. This allows us to fix the quark
model scale, $\mu_0$, such that upon evolution to 2~GeV the fraction
drops to $47\pm2$\%. The result of the LO DGLAP evolution is
$\mu_0=313^{+20}_{-10}$~MeV.  This perturbative renormalization scale
is unexpectedly and rather uncomfortably low. Yet, the prescription
has been favorably and independently confirmed by comparing to a
variety of other high-energy data or lattice calculations (see
\cite{Broniowski:2007si} and references therein). Moreover, the NLO
DGLAP modifications have been shown to yield moderate corrections as
well~\cite{Davidson:2001cc}. To summarize, {\em ``our approach =
chiral quark models + QCD evolution''}.

The pion $u$-quark tFFs, $B_{ni}^{\pi ,u}\left( t\right) $,
can be defined in a manifestly covariant way (see, e.g., \cite{Diehl:2010ru})
with the help of two auxiliary four-vectors, $a$ and $b$, satisfying $a^{2}=(ab)=0$ and $%
b^{2}\neq 0$. Then
\begin{eqnarray}
\langle \pi^{+}( p^{\prime }) | \overline{u}(0) i\sigma ^{\mu \nu }a_{\mu }b_{\nu }
\left( i \overleftrightarrow{D}a \right)^{n-1} u(0) | \pi^{+} (p) \rangle
=( a \cdot P )^{n-1} \frac{[ a \cdot p \, b \cdot p^{\prime}] }
{m_{\pi }}\sum_{\substack{ i=0,  \\ {\rm even}}}%
^{n-1}\left( 2\xi \right) ^{i}B_{ni}^{\pi ,u}\left( t\right) ,  \label{PionTme}
\end{eqnarray}
with the skewness parameter defined as $\xi =-{a \cdot q }/{( 2 a \cdot P ) }$
(we use the so-called symmetric kinematics).
The symbol $\overleftrightarrow{D}^{\beta }=\overleftrightarrow{\partial }^{\beta
}-igA^{\beta }$ denotes the QCD covariant derivative, and $\overleftrightarrow{%
\partial }^{\beta }=\frac{1}{2}\left( \overrightarrow{\partial }^{\beta }-%
\overleftarrow{\partial }^{\beta }\right) $. Further, $p^{\prime }$
and $p$ are the initial and final pion momenta, $P=\frac{1}{2}(p^{\prime
}+p) $, $q=p^{\prime }-p$, and $t=-q^{2}$. The factor $1/m_{\pi }$ is introduced by convention in order
to have dimensionless form factors.
Finally, the bracket $\left[ ...\right]$ denotes the antisymmetrization in the vectors $a$
and $b$. The tFFs defined in (\ref{PionTme}) apply to the $u$-quarks, while the tFFs for the $d$%
-quarks follow from the isospin symmetry, $B_{ni}^{\pi ,d}\left( t\right) =\left( -1\right) ^{n}B_{ni}^{\pi ,u}\left( t\right)$.
The definition of the corresponding tGPD is \cite{Belitsky:2005qn}
\begin{eqnarray}
\langle \pi ^{+}(p^{\prime })\mid \bar{u}(-a)i\sigma ^{\mu \nu }a_{\mu}b_{\nu }u(a)\mid \pi ^{+}(p)\rangle
=\frac{\left[  a \cdot p\,  b \cdot p^{\prime } \right] }
{m_{\pi }}\int_{-1}^{1}dX \,e^{-i X\, P \cdot a }E_{T}^{\pi, u}(X,\xi ,t), \label{PionTGPD}
\end{eqnarray}
where we do not write explicitly the gauge link operator needed to
keep the color gauge invariance.  The tFFs are related to the Mellin
moments of the tGPD,
\begin{equation}
\int_{-1}^{1}dX\,X^{n-1}E_{T}^{\pi, u}\left( X,\xi ,t\right)
=\sum_{\substack{ i=0, \\ {\rm even}}}^{n-1}\left( 2\xi \right)^{i}B_{ni}^{\pi ,u}\left( t\right),  \label{En}
\end{equation}
displaying the polynomiality property. Thus, the information carried
by tGPDs is contained in (infinitely many) tFFs. Some of them ($n=1,2$)
have been calculated on Euclidean lattices \cite{Brommel:2007xd}.

The details concerning the chiral quark models and parameters used
have been presented in~\cite{Broniowski:2010nt,Dorokhov:2011ew}. The
calculation of the tFFs and tGPDs is made through the use of standard
techniques at the one-quark-loop level, which corresponds to the
large-$N_c$ limit with confinement neglected.

\begin{figure*}
\centering
\includegraphics[width=0.48\textwidth]{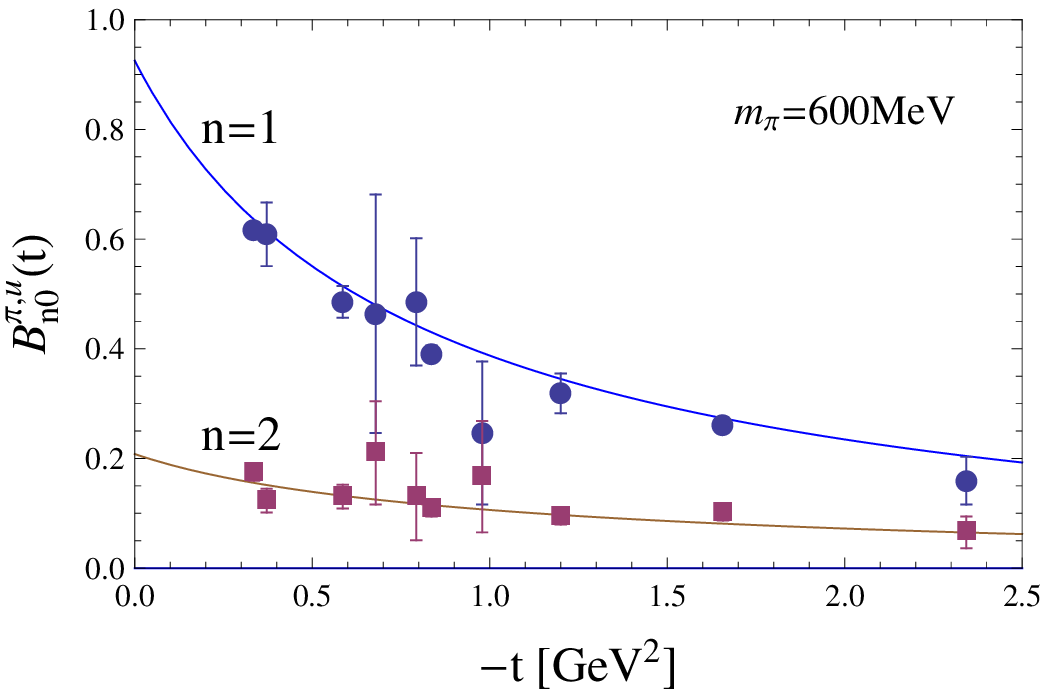} \includegraphics[width=0.48\textwidth]{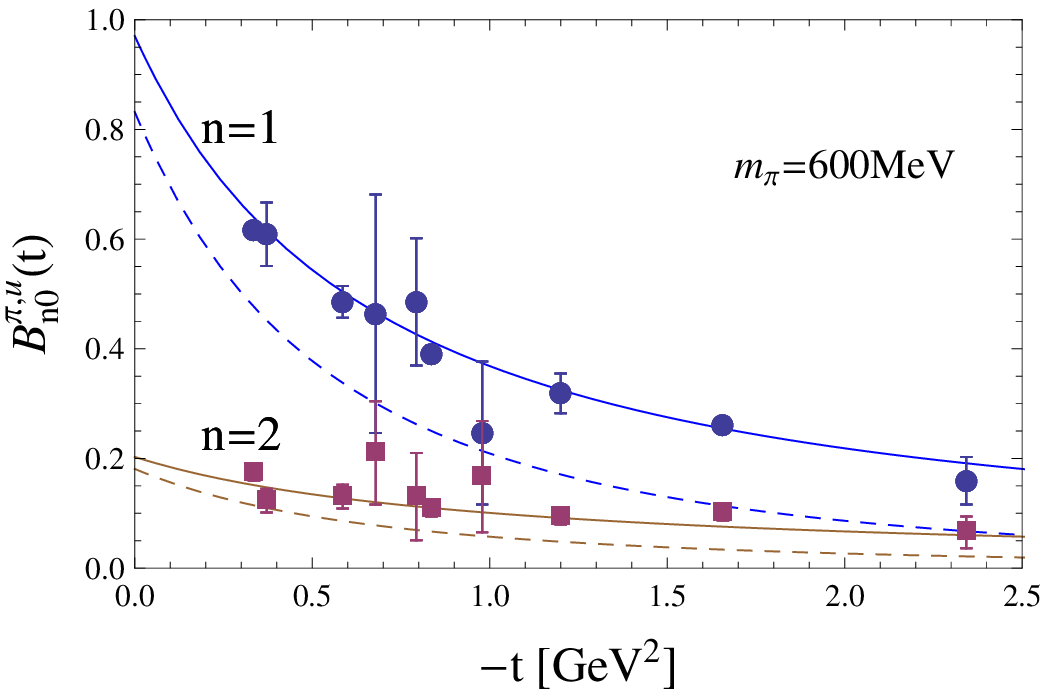}
\caption{The transversity form factors of the pion, $B^{\pi,u}_{10}(t)$ and $B^{\pi,u}_{20}(t)$, evaluated at $m_\pi=600$MeV in the local
NJL model (left panel) and in nonlocal models (right panel, solid line -- HTV model, dashed line -- instanton model)
and compared to the lattice data \cite{Brommel:2007xd}.  \label{fig:B12}}
\end{figure*}

Next, we describe the LO DGLAP-ERBL evolution. For the case of tFFs the procedure is very straightforward, as
it gives a triangular-matrix multiplicative structure (see, e.g.,~\cite{Broniowski:2009zh}).
Explicitly, with the
short-hand notation $B_{ni}=B_{ni}^{\pi }(t;\mu )$ and $B_{ni}^{0}=B_{ni}^{%
\pi }(t;\mu _{0})$, we have
\begin{eqnarray}
B_{10} &=&L_{1}B_{10}^{0},  \; B_{32} =\frac{1}{5}(L_{1}-L_{3})B_{10}^{0}+L_{3}B_{32}^{0},  \notag \\
&\dots &  \notag \\
B_{20} &=&L_{2}B_{20}^{0},  \; B_{42} =\frac{3}{7}(L_{2}-L_{4})B_{20}^{0}+L_{4}B_{42}^{0},  \notag \\
&\dots &  \notag \\
B_{30} &=&L_{3}B_{30}^{0},  \; B_{52} =\frac{2}{3}(L_{3}-L_{5})B_{30}^{0}+L_{5}B_{52}^{0},  \notag \\
&\dots &  \label{ev:ns}
\end{eqnarray}%
We define
\begin{equation}
L_{n}=\left( \frac{\alpha (\mu )}{\alpha (\mu _{0})}\right) ^{\gamma
_{n}^{T}/(2\beta _{0})}.
\end{equation}%
The anomalous dimensions in the transversity channel are given by
$\gamma _{n}^{T}=\frac{32}{3}\sum_{k=1}^{n}1/k-8$, $\beta _{0}=%
\frac{11}{3}N_{c}-\frac{2}{3}N_{f}$, and the running coupling constant is
$\alpha (\mu )={4\pi }/[{\beta _{0}\log (\mu ^{2}/\Lambda _{QCD}^{2})}]$,
with $\Lambda _{\mathrm{QCD}}=226~\mathrm{MeV}$ for $N_{c}=N_{f}=3$.

The two lowest tGFFs available from the lattice data, $B_{10}^{\pi,u}$ and $B_{20}^{\pi ,u}$, evolve multiplicatively as follows:
\begin{equation}
B_{n0}^{\pi ,u}(t;\mu )=B_{n0}^{\pi ,u}(t;\mu _{0})\left( \frac{\alpha
(\mu )}{\alpha (\mu _{0})}\right) ^{\gamma _{n}^{T}/(2\beta _{0})},
\end{equation}%
which numerically gives
\begin{eqnarray}
B_{10}^{\pi,u}(t;2~\mathrm{GeV})=0.75B_{10}^{\pi ,u}(t;\mu _{0}), \; B_{20}^{\pi ,u}(t;2~\mathrm{GeV})=0.43B_{20}^{\pi ,u}(t;\mu _{0}).
\label{evol:explicit}
\end{eqnarray}%
Note a stronger reduction for $B_{20}$ compared to $B_{10}$ as the result of the evolution.
In the chiral limit and at $t=0$
\begin{eqnarray}
B_{10}^{\pi,u}(t=0;\mu_{0})/m_{\pi}=\frac{N_{c} M}{4\pi^{2} f_{\pi}^{2}}, \; \;\;
\frac{B_{20}^{\pi,u}(t=0;\mu)}{B_{10}^{\pi,u}(t=0;\mu)}=\frac{1}{3}
\left( \frac{\alpha(\mu)}{\alpha(\mu_{0})}\right) ^{8/27}.  \label{LocLim2}
\end{eqnarray}

The results of the model calculation followed by evolution are presented in Fig.~\ref{fig:B12}. We note a very good agreement with the data
\cite{Brommel:2007xd} for the NJL model, as well as for the non-local HTV model. The agreement for the instanton model is worse, which shows that the
calculation can be used to discriminate between various approaches.

\begin{figure*}
\centering
\includegraphics[width=0.48\textwidth]{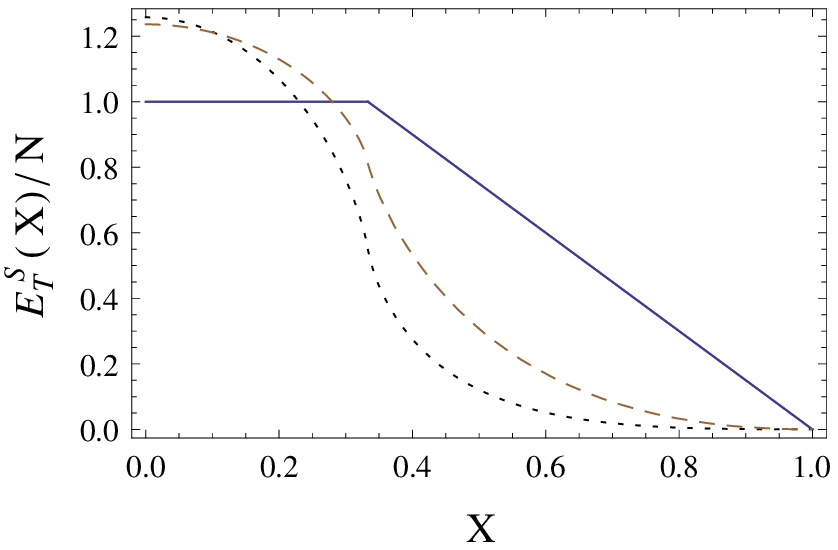} \includegraphics[width=0.48\textwidth]{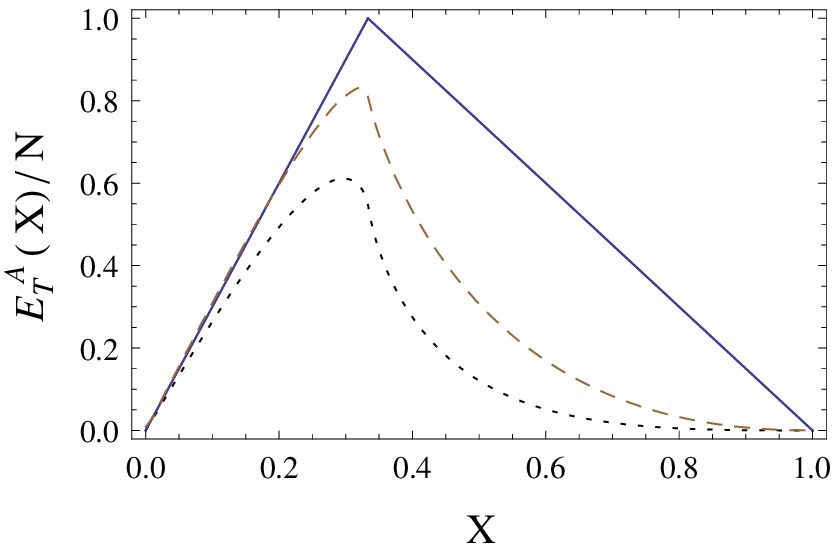}
\caption{The symmetric (left panel) and antisymmetric (right panel) tGPDs of the pion at $t=0$ and $\xi=1/3$, evaluated in the
NJL model in the chiral limit at the quark-model scale $\mu_0=313$~MeV (solid lines) and  evolved to the
scales $\mu=2$~GeV (dashed lines) and $1$~TeV (dotted lines). \label{fig:tGPD}}
\end{figure*}

\begin{figure*}
\centering
\includegraphics[width=0.48\textwidth]{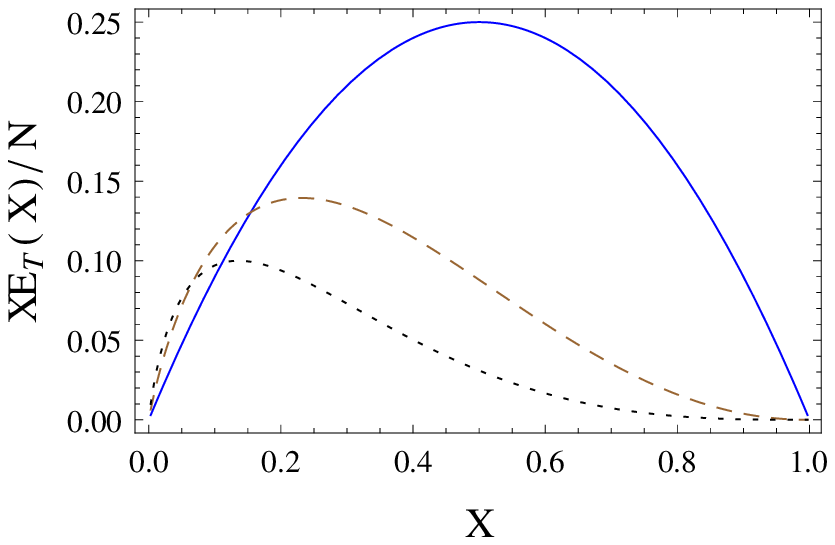} \includegraphics[width=0.48\textwidth]{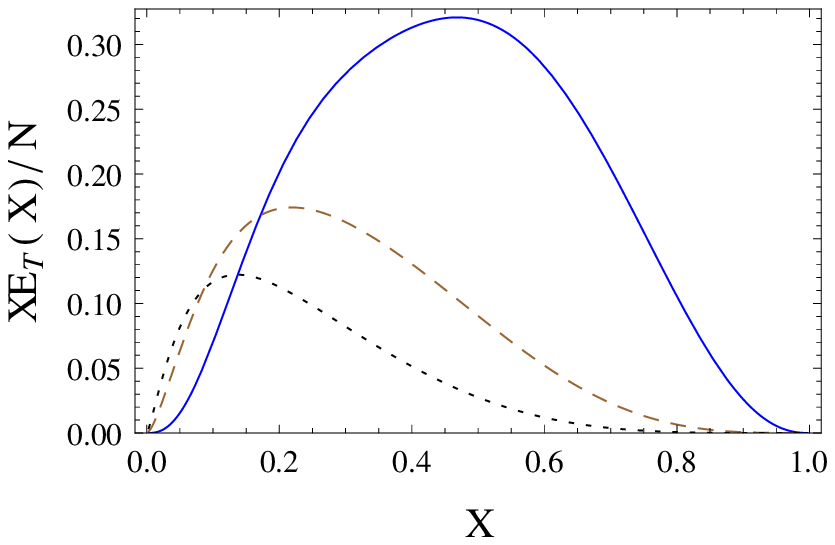}
\caption{The tGPD of the pion at $t=0$ and $\xi=0$, evaluated in the chiral limit in the
NJL model (left panel) and in the instanton model (right panel)
at the quark-model scale $\mu_0=313$~MeV (solid lines) and  evolved to the
scales $\mu=2$~GeV (dashed lines) and $1$~TeV (dotted lines). \label{fig:locnloc}}
\end{figure*}

Next, we show the results for the full tGPD. Having explored the $t$-dependence in tFFs, here we set $t=0$ and focus
on the $X$ behavior for fixed $\xi = 1/3$ or $\xi=0$. Since the evolution is different for the symmetric and antisymmetric
combinations, we explore the isovector and isoscalar tGPDs:
\begin{eqnarray}
&& E_{T}^{\pi ,I=1}\left( X,\xi ,Q^{2}\right) \equiv E_{T}^{\pi ,S}\left( X,\xi ,Q^{2}\right) = E_{T}^{\pi }\left(
X,\xi ,Q^{2}\right) +E_{T}^{\pi }\left( -X,\xi,Q^{2}\right) ,   \nonumber \\
&& E_{T}^{\pi ,I=0}\left( X,\xi ,Q^{2}\right) \equiv E_{T}^{\pi ,A}\left( X,\xi ,Q^{2}\right) \nonumber = E_{T}^{\pi }\left(
X,\xi ,Q^{2}\right) -E_{T}^{\pi }\left( -X,\xi,Q^{2}\right) .  \label{ETnI01}
\end{eqnarray}
The evolution has been carried out with the method involving the Gegenbauer moments\cite%
{Kivel:1999sk,Kivel:1999wa,Manashov:2005xp,Kirch:2005tt}. The results for $\xi=1/3$ in the NJL model are shown
in Fig.~\ref{fig:tGPD}. The curves are conventionally normalized in such a way that at the quark model scale
\begin{equation}
\int_{0}^{1}dXE_{T}^{\pi ,S}(X,\xi ,t=0;\mu _{0})/N=\frac{1+\xi }{2}.
\end{equation}
We note the decreasing, shifting to lower $X$, effect of the QCD evolution. In Fig.~\ref{fig:locnloc} we compare the
result in the NJL model (left panel) and the nonlocal instanton model (right panel). Except for different
end-point behavior (see \cite{Dorokhov:2011ew} for details), the results are qualitatively very similar.

In conclusion we wish to stress that the
absolute predictions for the multiplicatively evolved $B_{10}$ and $B_{20}$ agree
remarkably well with the lattice results, supporting the assumptions of numerous calculations
following the same ``chiral-quark-model + QCD evolution'' scheme.
It would also be interesting to access the tGPDs directly through the use of the transverse-lattice techniques \cite{Burkardt:2001jg,Dalley:2001gj,Dalley:2004rq,Dalley:2002nj}.



\end{document}